\documentclass[12pt,preprint]{aastex}

\shorttitle{Hot core and outflows in G34.43+00.24 MM3}
\shortauthors{Sakai et al.}

\begin{document}


\title{ALMA OBSERVATIONS OF THE IRDC CLUMP G34.43+00.24 MM3: HOT CORE AND MOLECULAR OUTFLOWS}


\author{Takeshi Sakai\altaffilmark{1}, Nami Sakai\altaffilmark{2}, Jonathan B. Foster\altaffilmark{3}, Patricio Sanhueza\altaffilmark{4}, James M. Jackson\altaffilmark{4}, Marc Kassis\altaffilmark{5},
Kenji Furuya\altaffilmark{6}, Yuri Aikawa\altaffilmark{6}, Tomoya Hirota\altaffilmark{7} and Satoshi Yamamoto\altaffilmark{2}}

\altaffiltext{1}{Graduate School of Informatics and Engineering, The University of 
Electro-Communications, Chofu, Tokyo 182-8585, Japan.}
\altaffiltext{2}{Department of Physics, Graduate School of Science, The University of Tokyo, Tokyo 113-0033, Japan.}
\altaffiltext{3}{Yale Center for Astronomy and Astrophysics, Yale University, New Haven, CT 06520, USA.}
\altaffiltext{4}{Institute for Astrophysical Research, Boston University, Boston, MA 02215, USA.}
\altaffiltext{5}{W. M. Keck Observatory, 65-1120 Mamalahoa Hwy., Kamuela, HI 96743, USA.}
\altaffiltext{6}{Department of Earth and Planetary Sciences, Kobe University, Kobe 657-8501, Japan.}
\altaffiltext{7}{National Astronomical Observatory of Japan, Osawa, Mitaka, Tokyo 181-8588, Japan.}


\begin{abstract}
We have observed a cluster forming clump (MM3) associated with the infrared dark cloud G34.43+00.24 in the 1.3 mm continuum and the CH$_3$OH, CS, $^{13}$CS, SiO, CH$_3$CH$_2$CN, and HCOOCH$_3$ lines with the Atacama Large Millimeter/submillimeter Array and in $K$-band with the Keck telescope.
We have found a young outflow toward the center of this clump in the SiO, CS, and CH$_3$OH lines. 
This outflow is likely driven by a protostar embedded in a hot core, which is traced by the CH$_3$CH$_2$CN, HCOOCH$_3$, $^{13}$CS, and high excitation CH$_3$OH lines.
The size of the hot core is about 800$\times$300 AU in spite of its low mass ($<$1.1 $M_{\odot}$), suggesting a high accretion rate or the presence of multiple star system harboring a few hot corinos.
The outflow is highly collimated, and the dynamical timescale is estimated to be less than 740 yr.
In addition, we have also detected extended emission of SiO, CS, and CH$_3$OH, which is not associated with the hot core and the outflow. This emission may be related to past star formation activity in the clump. 
Although G34.43+00.24 MM3 is surrounded by a dark feature in infrared, it has already experienced active formation of low-mass stars in an early stage of clump evolution.

\end{abstract}





\section{Introduction}

It is known that 70-90 $\%$ of stars are born in clusters in giant molecular clouds (Lada \& Lada, 2003). In particular, high-mass stars are usually born as a member of a cluster.
Thus, understanding the formation mechanism of clusters is one of the most important issues for astronomy.
A molecular clump (size$>$0.1 pc, mass$>$a few times 10 $M_{\odot}$) is thought to be a birthplace of clusters.
Based on observations of molecular clumps, we can address the following questions: how do clumps fragment into small cores, how and when is the mass of each cluster member determined, and how does star formation activity affect fragmentation?
In order to answer these questions, the exploration of early evolutionary stages of clumps is crucial.
Infrared dark clouds (IRDCs) are thought to be good target, for this purpose.

IRDCs are dark in the mid-infrared band, and are distributed throughout the galactic plane (Jackson et al. 2008). Many clumps associated with IRDCs have been found (e.g. Rathborne et al. 2006). They are cold (10--20 K) and dense ($\sim$10$^{4-5}$ cm$^{-3}$), and are thought to host very early stages of cluster (and high-mass star) formation.
In addition, hot molecular cores with temperatures of 100 K or higher are associated with some IRDCs (e.g. Rathborne et al. 2011). 
These represent the early stages of star formation.

G34.43+00.24 is a filamentary IRDC (Garay et al. 2004; Rathborne et al. 2005; 2006). 
The distance to this IRDC was determined to be 1.56$^{+0.12}_{-0.11}$ kpc by VLBI maser parallax observations (Kurayama et al. 2011).
Rathborne et al. (2006) mapped this IRDC at 1.2 mm continuum and revealed that G34.43+00.24 consists of 9 clumps, as shown in Figure 1a.
Among them, MM2 contains an ultracompact HII region (Miralles et al. 1994). This means that high-mass star formation is taking place in MM2.

G34.43+00.24 MM3 is less luminous in the infrared band than MM1 or MM2.
As seen in Figure 1a, MM3 is surrounded by a dark feature at 24 $\mu$m.
Although star formation has already started in MM3 (Rathborne et al. 2005), MM3 is thought to be in an earlier evolutionary stage than MM1 and MM2 (Garay et al. 2004).
The mass of MM3 clump was originally reported to be 171 $M_{\odot}$ using the kinematic distance of 3.7 kpc (Rathborne et al. 2010). 
However, there is a large difference between the parallax distance and other methods (i.e. the kinematic distance and near-infrared extinction distances) for determining the distance to this cloud (Foster et al. 2012).
In this Letter, we adopt the maser parallax distance because its uncertainty is tighter than those of the other methods.
Using the new distance of 1.56 kpc gives a mass of 30 $M_{\odot}$.

Sanhueza et al. (2010) reported an outflow in CO $J$=3-2 toward MM3. 
In addition, the CH$_3$OH and SiO line widths are much broader than the N$_2$H$^+$, H$^{13}$CO$^+$, and HN$^{13}$C line widths in MM3 (Sakai et al. 2008, 2010, Sanhueza et al. 2012). Since the CH$_3$OH and SiO lines trace shocked regions, these results indicate strong interactions between outflow and dense gas. Furthermore, the narrow N$_2$H$^+$ line suggests that a substantial amount of cold and quiescent gas still remains in MM3.
Thus, MM3 is thought to be a good target to investigate an early stage of cluster formation. 

However, details of star formation activity within MM3 are still incomplete as previous single-dish observations lacked sufficient angular resolution.
We conducted sub-arcsecond resolution observations in the continuum and the CH$_3$OH, CS, $^{13}$CS, SiO, CH$_3$CH$_2$CN, and HCOOCH$_3$ lines at 1.3 mm toward this source for the first time with ALMA in its early science operation, and also in the $K$-band with the Keck telescope. From the Keck observations, we investigated the distribution of shocked H$_2$ gas and the population of young stellar objects with much higher angular resolution (0.15$^{\prime\prime}$) than that of the previous $Spitzer$ data ($>2^{\prime\prime}$).
In this Letter, we show the ALMA and Keck data, and discuss star formation activity in G34.43+00.24 MM3.

\section{OBSERVATIONS}

\subsection{ALMA}

We observed G34.43+00.24 MM3 with ALMA Band 6 on 2012 August 11, 15, and 26.
The phase center was (R.A.(J2000), Dec.(J2000)) = (18$^{\rm h}$53$^{\rm m}$20.4$^{\rm s}$, 1$^{\circ}$28$^{\prime}$23.0$^{\prime\prime}$).
The observations were carried out with the extended configuration using 23-26 antennas, whose baseline coverage is from 20.6 m to 399.6 m.
The spectrometers were used with the 234 MHz mode with a 61 kHz channel width.
Since we averaged 8 channels for all the data in order to reduce the noise, the channel width shown in this Letter is 488 kHz, corresponding to the velocity width of  0.64 km s$^{-1}$ at 230 GHz.

The bandpass calibrations were carried out by observing J1924-292.
The flux calibrations were done by observing Neptune and J1751+096.
Typical system noise temperature was 100-150 K.
The data were reduced by the CASA software package.
The 1.3 mm (236.930 GHz) continuum image was made by averaging the line-free channels of the four 234 MHz basebands between 228-245 GHz.
The line images were obtained by CLEANing the dirty images after subtracting the continuum directly from the visibilities.
The observed lines and their parameters, including synthesized beam size for each line, are listed in Table 1.

\subsection{Keck}

We observed G34.43$+$00.24 MM3 on 2010 August 20. We observed in the $K_p$ (2.124 $\mu$m) filter using NIRC2 on Keck II with laser-guide star adaptive optics. The observations used the wide camera (40 mas/pixel and 40$\times$40 arcsecond field of view). To avoid saturating on the brightest sources we took short (3 second) exposures with 30 co-adds. We used a five-point dither pattern with 5 arcsecond offsets between exposures to allow for sky-subtraction. Total exposure in $K_p$ was 1 hour. The data were reduced in a standard fashion, including sky-subtraction from a running median of five individual exposures and applying a correction for the known distortion of NIRC2. Cross-matching with the 2MASS catalog provided the final coordinate registration and photometric calibration. Due to very strong seeing fluctuations during these observations, the effective resolution of these images was only 0.15 arcseconds. Further details and analysis of these observations will be presented in Foster et al. (in preparation).

\section{RESULTS}

Figure 1b shows the continuum image toward G34.43+00.24 MM3 at 1.3 mm.
The strongest peak is seen toward the northeast of the image (hereafter we call this peak as Peak A).
Peak A is apparently distinct from the $Spitzer$ sources reported in Shepherd et al. (2007), and is recognized by this observation for the first time. 
The absence of $K$-band emission (Figure 2) suggests that this is a deeply embedded source.
The peak intensity of this source is 14 mJy beam$^{-1}$, and its integrated flux density is 38 mJy.
The deconvolved source size is $1.2^{\prime\prime} \times 1.2^{\prime\prime}$ with P.A. = -77.9$^{\circ}$, corresponding to 1900 AU $\times$ 1900 AU at the distance of 1.56 kpc.
On the other hand, no continuum source is found toward the $Spitzer$ sources.
This implies that the $Spitzer$ sources are evolved, and may have already dissipated their parent dense cores.

The second continuum peak (Peak B) is seen south of the strongest peak.  
Molecular line emission is not clearly seen at this peak, and hence this continuum source might not be associated with the IRDC clump.  We will not further discuss the properties of this continuum source in this Letter.

The flux density integrated over the field of view ($\sim$27$^{\prime\prime}$) is 55 mJy.
Rathborne et al. (2006) reported the 1.2 mm continuum emission integrated over a diameter of 24$^{\prime\prime}$ to be 1.02 Jy. Thus, most of the flux is resolved out in the ALMA observation.

Toward Peak A, compact CH$_3$OH $J_K$ = $10_2$--$9_3$ $A^-$ emission was detected, as shown in Figure 1c.
Since the upper state energy of this CH$_3$OH line is as high as 165 K, its detection means that Peak A harbors a hot molecular core.
In addition, complex organic molecules, CH$_3$CH$_2$CN (Figure 1d) and HCOOCH$_3$ (Figure 1e), are detected toward this source.
Since these complex organic molecules are specific to hot cores and hot corinos (e.g. Herbst \& van Dishoeck, 2009), Peak A most likely contains a hot core unrevealed in all previous observations.

The integrated intensity distribution of CS (Figure 1f) has two intense peaks toward the northern and southern positions of Peak A. 
The CS emission looks as if it traces the collimated outflow from a protostar associated with Peak A.  
The SiO emission (Figure 1h), which is a good shock tracer, is also strong toward the CS peaks.
Thus, the CS peaks are thought to trace shocked regions due to interaction between the outflow and dense gas.  This view is consistent with that proposed by Sakai et al. (2010) on the basis of their single-dish observations (see Figure 9 in Sakai et al.).
In contrast, the optically thinner $^{13}$CS emission is strong toward Peak A (Figure 1g).
This indicates that CS is more abundant in the hot core than in the shocked regions, and that the CS emission is highly optically thick and substantially resolved out. 

In addition to the two peaks, the CS emission is extended toward the southwestern part from the continuum source (Peak A). Although a few local peaks are found, no continuum sources are associated with them.
In the CH$_3$OH $J_K$ = $5_0$--$4_0$ $A^+$ image (Figure 1i), this feature is more prominent than any other emission located near Peak A.
On the other hand, the SiO emission tends to be localized around Peak A, except that a weak SiO peak is found at the most southwestern peak of the extended CS emission.

In the $K$-band image (Figure 2), we can see the diffuse emission toward the south lobe of the CS outflow.  
This emission is due to the H$_2$ line at 2.12 $\mu$m.
Chambers et al. (2009) reported a "green fuzzy" feature in this region, which is a 4.5 $\micron$ excess due to the shock excited H$_2$ or CO lines.
Our results clearly support that the 4.5 $\mu$m excess is due to the interaction between the outflow and dense gas.
A similar diffuse feature in the $K$-band is not observed toward the north lobe of the outflow, perhaps because it is absorbed by the foreground dense gas.
There is a $K$-band source associated with the southwestern CS emission. This source is also associated with the $Spitzer$ source \#27 reported in Shepherd et al. (2007).  The extended $K$-band emission from this source arises from H$_2$ line, although its extent does not match the CS emission well.

\section{ANALYSIS AND DISCUSSION} 

\subsection{Properties of the Outflow}

Figure 3 shows the position velocity (PV) diagrams of SiO, CS and CH$_3$OH along a straight line through Peak A (Figure 2).
The SiO profiles show broad emission toward the northern and southern peaks.
This velocity structure is consistent with the bow-shocked outflow model (Lee et al. 2001).
The velocity width is as broad as $\sim$20 km s$^{-1}$ at the two peaks which are symmetrically displaced by $\sim$2$^{\prime\prime}$ from Peak A. The broad emission is also seen in CS (Figure 3b). In the CS map, a linear velocity gradient is visible, as shown by the yellow dotted line.
Such a linear velocity structure has been reported toward some sources (e.g. HH211; Gueth \& Guilloteau 1999), and is explained by the wide-angle outflow model, where the emission comes from the cavity wall of the outflow (Lee et al. 2001).
Thus, the velocity structure of the CS emission can be interpreted in terms of a wide-angle outflow with a bow shock.
In addition, the velocity structure indicates that the outflow axis is close to perpendicular to the line of sight (see Figure 22 of Lee et al. 2000). 

Toward Peak A, weak SiO emission, as well as the CH$_3$OH $J_K$ = $10_2$--$9_3$ $A^-$ emission, is seen around the velocity of 60 km s$^{-1}$.
At the same time, weak emission is seen in the velocity range of 25-50 km s$^{-1}$ toward Peak A. This emission may be the CH$_3$CH$_2$CN 29$_{5,25}$--28$_{5,24}$ line.
In the red contour of Figure 3, we also see the CH$_3$CH$_2$CN emission, 27$_{0,27}$--26$_{1,26}$ and 24$_{2,23}$--23$_{1,22}$, near the velocity of 20 km s$^{-1}$.

In the CH$_3$OH map (Figure 3c), the broad components are observed at the two shocked regions. However, the blue- and redshifted velocity components seem to be located in an opposite direction in comparison with the SiO and CS cases. 
Furthermore, the broad components of CH$_3$OH are seen farther away from Peak A than the CS and SiO peaks.
A similar CH$_3$OH distribution was reported in the NGC2264 region (Saruwatari et al. 2011).
The different distributions of SiO, CS, and CH$_3$OH mean chemical differentiation in the shock. 

The outflow parameters are derived by use of the following equations; $R_{\rm outflow}$ = $R_{\rm obs}$/$\sin i$, $V_{\rm outflow}$ = $V_{\rm max}$/$\cos i$, and $t_{\rm dyn}$ = $R_{\rm outflow}$/$V_{\rm outflow}$, where $R_{\rm obs}$ is a projected size, $V_{\rm max}$ is the maximum velocity of the outflow in the PV map, and $i$ is an inclination angle.
By assuming the inclination angle of 45 degree for simplicity, the outflow size, outflow velocity, and dynamical timescale are evaluated to be $\sim$4400 AU, $\sim$28 km s$^{-1}$, and $\sim$740 yr, respectively, from the SiO data.
Note that we obtain almost same results from the CS data.
The dynamical timescale of this outflow is shorter than those typically found in high-mass protostellar objects (5$\times$10$^3$-1.8$\times$10$^5$ yr; Beuther et al. 2002) and intermediate-mass hot cores (1.2$\times$10$^3$-4.0$\times$10$^3$ yr; Sanchez-Monge et al. 2010 and references therein).
Thus, this outflow is likely very young.

\subsection{Properties of the Hot Core}

Using the observed intensities of the high excitation CH$_3$OH lines ($J_K$=$10_2$--$9_3$ $A^-$ and $J_K$=$5_0$--$4_0$ $A^+$) toward Peak A, we calculated the rotation temperature of CH$_3$OH to be 134$\pm$10 K under assumptions of local thermodynamic equilibrium condition and optically thin emission.
Hence, Peak A definitively contains a hot core, and the CH$_3$OH $J_K$=$10_2$--$9_3$ $A^-$ emission likely traces a hot ($>$100 K) region.
The size of the CH$_3$OH $J_K$=$10_2$--$9_3$ $A^-$ source (0.97$\pm$0.12$^{\prime\prime}$ $\times$ 0.69$\pm$0.09$^{\prime\prime}$) is slightly larger than the synthesized beam, although it has a relatively large uncertainty.
Its deconvolved size is 0.5$^{\prime\prime}$ $\times$  0.2$^{\prime\prime}$ with P.A. = -56.9$^\circ$, corresponding to 800 AU $\times$ 300 AU.
This size is larger than those of the low-mass hot corinos in IRAS 16293-2422 ($\sim$80 AU; Crimier et al. 2010) and other low-mass cores (13-53 AU; Maret et al. 2004).

A larger size of a hot ($>$100 K) region means a larger luminosity of an embedded protostar compared to low-mass hot corinos.
The luminosity of a protostar depends on its age and accretion rate.
The dynamical timescale of the outflow is about 7-9 times shorter than that of IRAS 16293-2422 (5-7$\times$10$^3$ yr; Hirano et al. 2001).
If the size of the hot core is larger than the low-mass hot corinos, the short dynamical time scale may suggest a higher accretion rate in the G34.43+00.24 MM3 hot core.
Alternatively, this source may be a multiple star system harboring a few hot corinos.
To investigate these possibilities, higher angular resolution observations are necessary.

Since the size of the 1.3 mm continuum emission at Peak A is larger than the size of the hot core, the 1.3 mm continuum emission is likely to include a contribution from a relatively cold envelope.
We derive the mass of the continuum source by using the same method as reported in Rathborne et al. (2006).
Rathborne et al. (2010) reported the dust temperature of this clump to be 29 K by fitting the SED at the infrared and radio wavelengths.
The temperature is expected to be in the range between the two estimates of the temperature (29--134 K).
In this temperature range, the mass is derived to be 0.2-1.1 $M_\odot$.
Despite the problem of the missing flux, it is unlikely that a high-mass star will be born in this source.
Since we have no definitive evidence of high-mass star formation in this clump, 
we suggest that it may be the formation site of a low- and intermediate-mass star cluster.
Nevertheless, a very young protostar is growing in Peak A, and a hot core is associated with it.
This source is important to investigate the very early stage of star formation in a cluster forming region, which deserves further detailed observations for physical and chemical characterizations.

\subsection{Origin of the Southwestern Molecular Emission}

Since there are no luminous continuum sources around the southwestern-extended emission, high- or intermediate-mass protostars may not be associated with them. 
In Figure 4, we show the distributions of the redshifted and blueshifted components of CH$_3$OH.
The blueshifted emission is seen near the $Spitzer$ source \#27. This emission may be related to the outflow from \#27.  However, the redshifted emission of the outflow is not seen near \#27. Instead, it is seen toward the east of Peak A.
Since the redshifted emission is too far from \#27, it may not be the outflow driven by \#27, but probably by the $Spitzer$ source \#29.
We found several $K$-band sources around \#29, but it is unclear which $K$-band source drives this redshifted outflow.
Several low mass stars might contribute to the southwestern-extended emission.

The existence of the outflow emission and the extended H$_2$ emission indicate star formation activities other than that of Peak A.
In Figure 4, we can also see redshifted emission toward the north part of the image.
This emission could be related to the past star formation activity in this clump.
Thus, it is likely that several low-mass stars have already formed before the protostar was born in Peak A.
Some $K$-band sources display evidence that they are low-mass young stellar objects, whose properties will be discussed in the forthcoming paper (Foster et al., in preparation).

Our results suggest that low-mass stars have been formed in a very early evolutionary stage of the G34.43+00.24 MM3 clump.
Sakai et al. (2012) reported that the deuterium fractionation is low in this source.
This may be due to heating by the past star formation activity in this clump.
We will discuss the deuterium fractionation ratio in this clump separately (Sakai et al., in preparation).
To investigate whether low-mass stars are generally formed in very early stage of clump formation, statistical studies for many IRDCs are necessary.

\acknowledgments

This Letter makes use of the following ALMA data: ADS/JAO.ALMA\#2011.0.00656.S. ALMA is a partnership of ESO (representing its member states), NSF (USA) and NINS (Japan), together with NRC (Canada) and NSC and ASIAA (Taiwan), in cooperation with the Republic of Chile. The Joint ALMA Observatory is operated by ESO, AUI/NRAO, and NAOJ. We are grateful to the ALMA staffs. Some of the data presented herein were obtained at the W.M. Keck Observatory, which is operated as a scientific partnership among the California Institute of Technology, the University of California and the National Aeronautics and Space Administration. The Observatory was made possible by the generous financial support of the W.M. Keck Foundation. We would like to thank the observatory director, Dr. Taft Armandroff, who graciously provided his time for the acquisition of this data.
KF is supported by the Research Fellowship from the Japan Society for the Promotion of Science (JSPS) for Young Scientists.
This study is supported by Grant-in-Aid from Ministry of Education, Culture, Sports, Science, and Technologies of Japan (21224002, 23740146, 25400225 and 25108005).

\clearpage



\begin{figure}
\epsscale{1.0}
\plotone{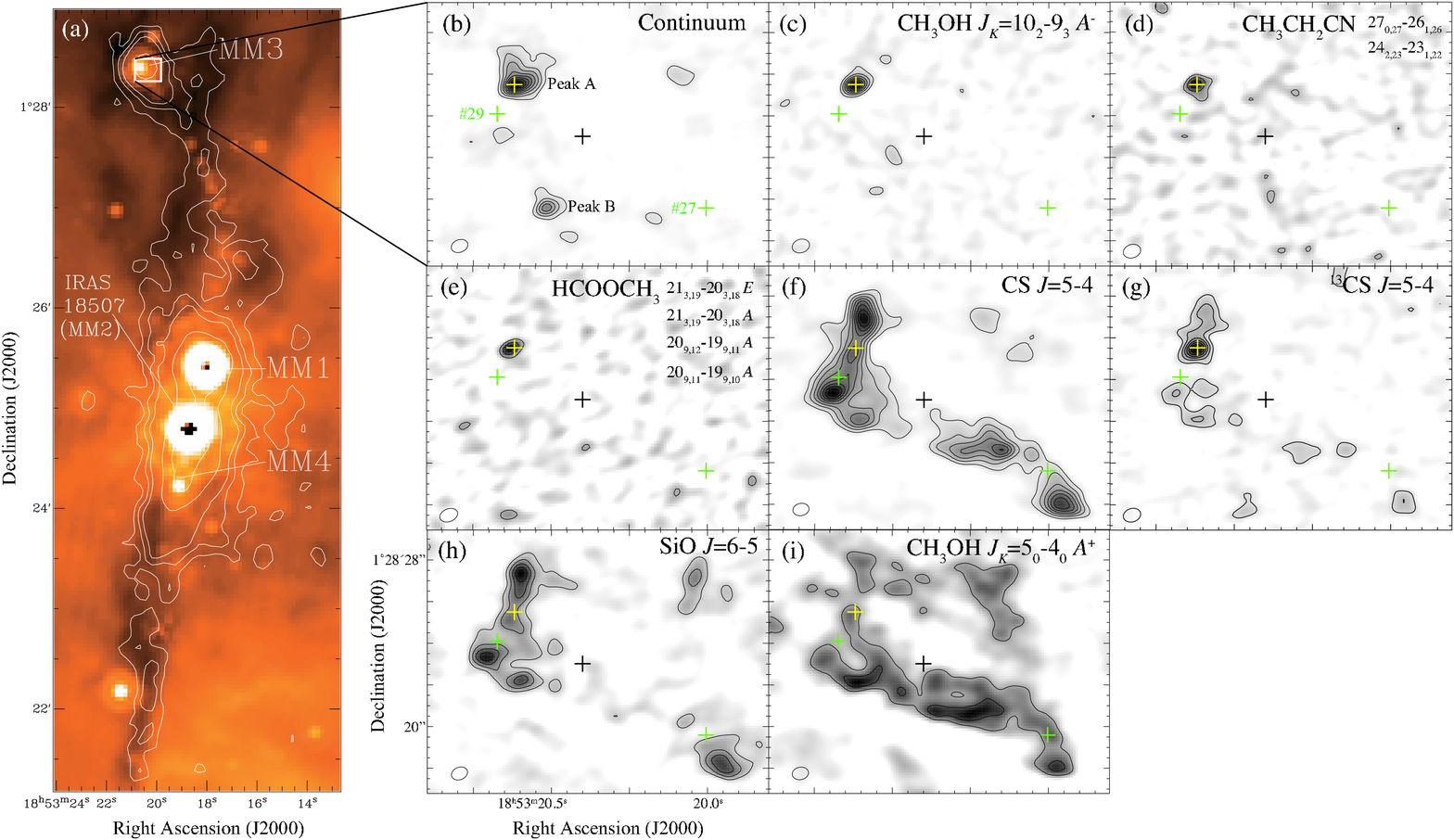}
\caption{(a) $Spitzer$ 24 $\mu$m color image overlaid with the 1.2 mm continuum (Rathborne et al. 2006). (b) 1.3 mm continuum emission from ALMA, (c) CH$_3$OH $J$=10$_2$--9$_3$ $A^-$, (d) CH$_3$CH$_2$CN $27_{0,27}$--$26_{1,26}$ and $24_{2,23}$--$23_{1,22}$, (e) HCOOCH$_3$ $21_{3,19}$--$20_{3,18}$ $E$, $21_{3,19}$--$20_{3,18}$ $A$, $20_{9,12}$--$19_{9,11}$ $A$, and $20_{9,10}$--$19_{9,10}$ $A$, (f) CS $J$=5--4, (g) $^{13}$CS $J$=5--4, (h) SiO $J$=6--5, (i) CH$_3$OH $J$=5$_0$--4$_0$ $A^+$.
Contour levels start and increase in steps of 3$\sigma$ [(b) $3\sigma=1.4$ mJy beam$^{-1}$, (c) $3\sigma=60$ mJy beam$^{-1}$ km s$^{-1}$, (d) $3\sigma=50$ mJy beam$^{-1}$ km s$^{-1}$, (e) $3\sigma=70$ mJy beam$^{-1}$ km s$^{-1}$, (f) $3\sigma=480$ mJy beam$^{-1}$ km s$^{-1}$, (g) $3\sigma=60$ mJy beam$^{-1}$ km s$^{-1}$, (h) $3\sigma=290$ mJy beam$^{-1}$ km s$^{-1}$, (i) $3\sigma=690$ mJy beam$^{-1}$ km s$^{-1}$]. The black, yellow and green cross marks represent the position of the phase center, the peak of CH$_3$OH $J$=10$_2$--9$_3$ $A^-$ and the $Spitzer$ sources, respectively. The name of the $Spitzer$ sources are referred from Shepherd et al. (2007). \label{fig1}}
\end{figure}

\clearpage


\begin{figure}
\epsscale{1.0}
\plotone{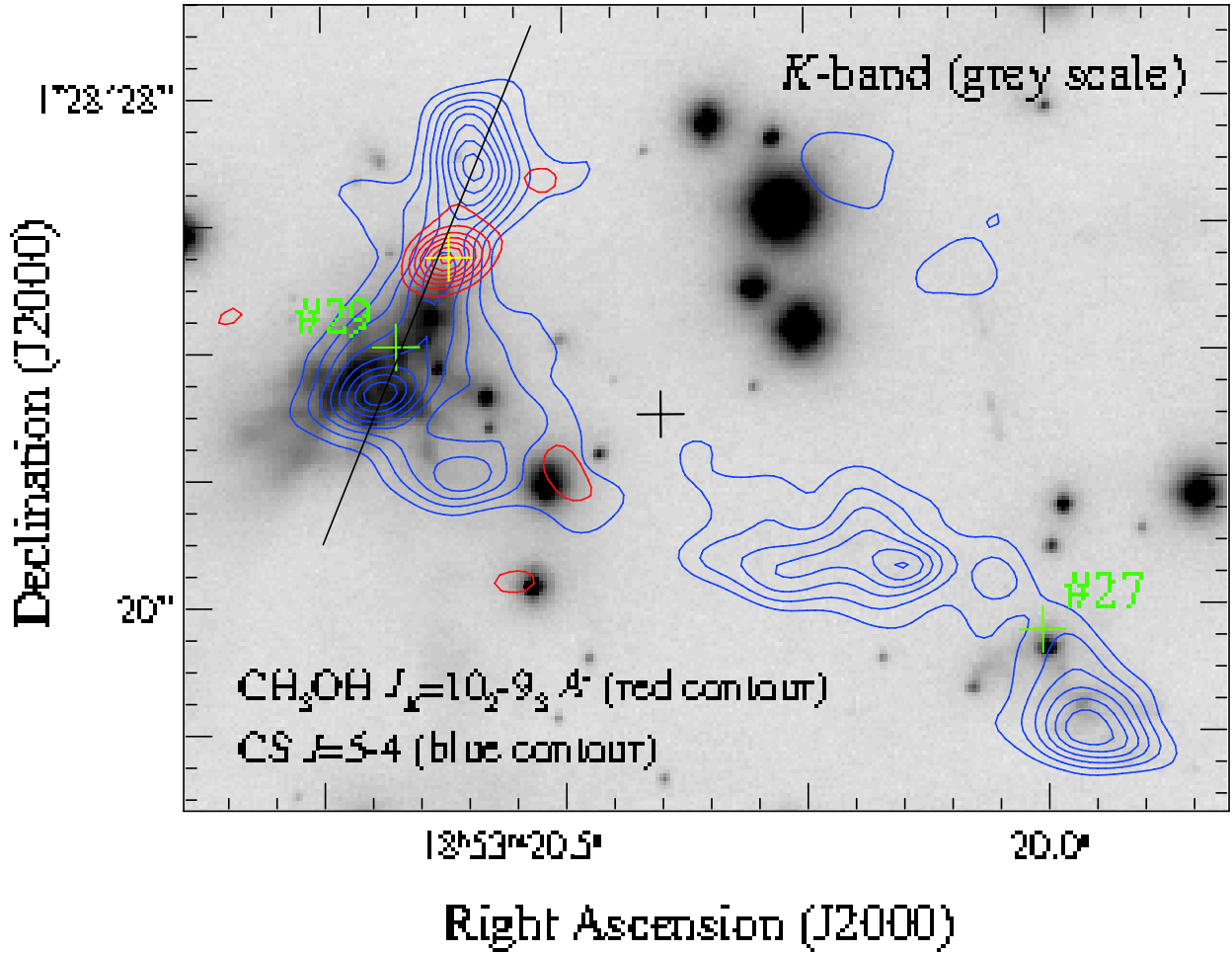}
\caption{CS $J$=5--4 and CH$_3$OH $J_K$=10$_2$--9$_3$ $A^-$ contour images superposed to the $K$-band image from the Keck telescope. The solid line is for the position velocity diagram in Figure 3. \label{fig2}}
\end{figure}

\clearpage

\begin{figure}
\epsscale{.40}
\plotone{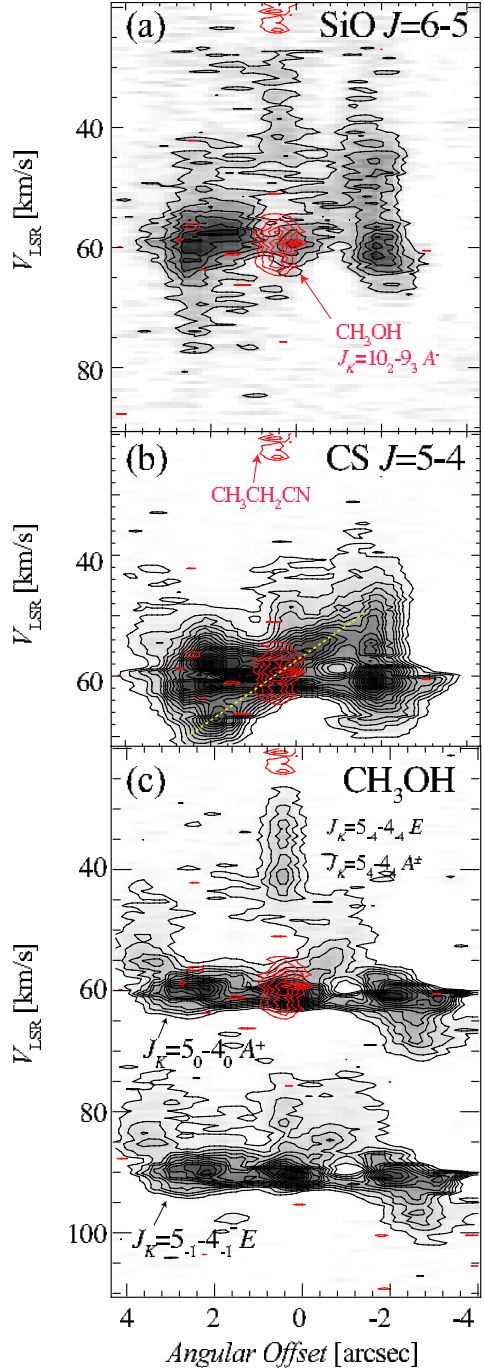}
\caption{Position velocity diagrams of SiO (a), CS (b), and CH$_3$OH (c) through the line indicated in Figure 2. CH$_3$OH $J_K$=$10_2$--$9_3$ $A^-$ and CH$_3$CH$_2$CN ($27_{0,27}$--$26_{1,26}$ and $24_{2,23}$--$23_{1,22}$) are overlaid in red. The lowest contour level and the contour step are (12 mJy beam$^{-1}$, 12 mJy beam$^{-1}$), (12 mJy beam$^{-1}$, 20 mJy beam$^{-1}$), and (10 mJy beam$^{-1}$, 17 mJy beam$^{-1}$) for (a), (b) and (c), respectively. The yellow dotted line in (b) indicates the linear velocity structure of the outflow.\label{fig3}}
\end{figure}


\clearpage

\begin{figure}
\epsscale{1.0}
\plotone{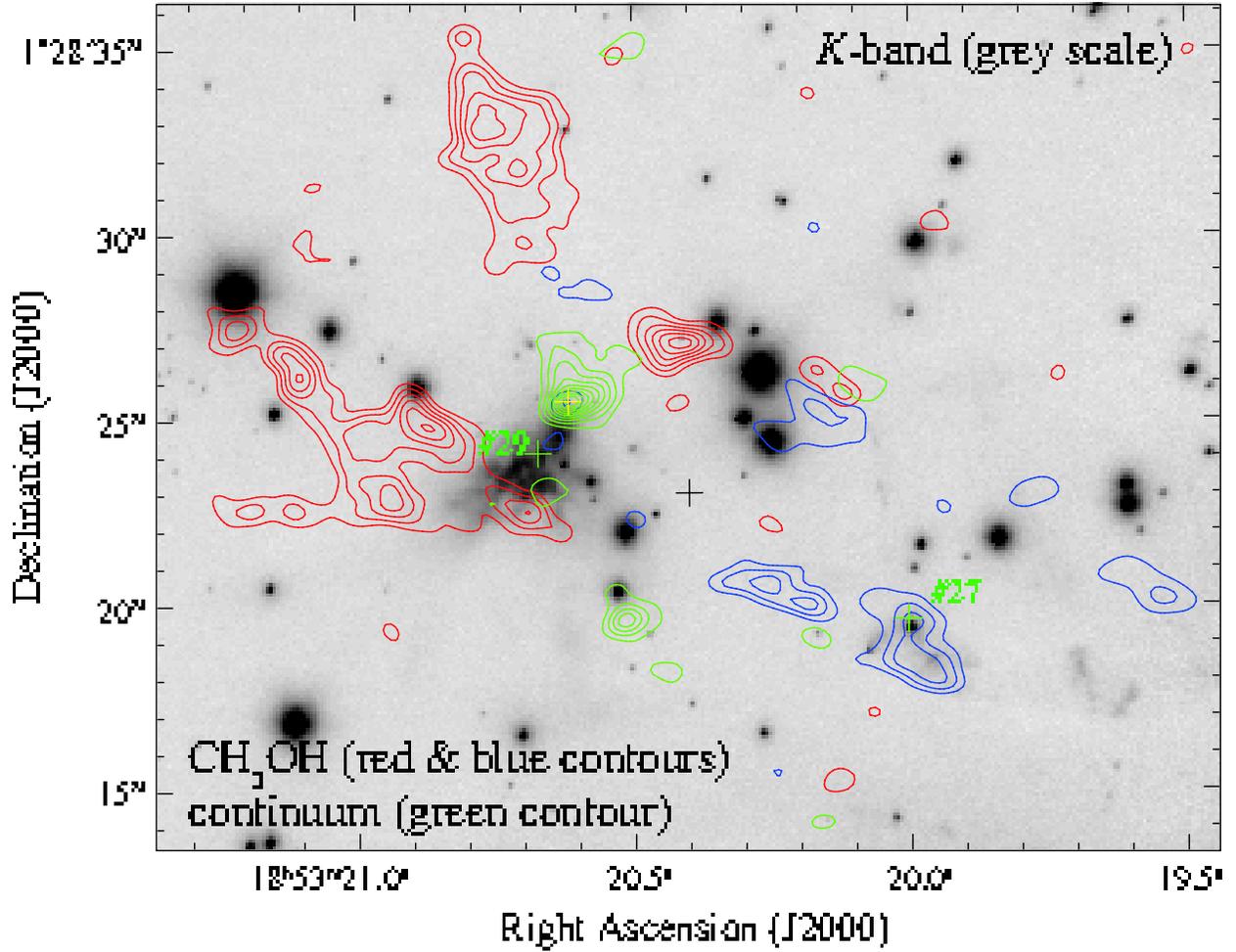}
\caption{Integrated intensity map of the redshifted velocity component (66--81 km s$^{-1}$) of CH$_3$OH $J_K$=$5_{-1}$--$4_{-1}$ $E$ (red) and the blueshifted velocity component (39--55 km s$^{-1}$) of CH$_3$OH $J_K$=$5_{0}$--$4_{0}$ $A^+$(blue) superposed to the $K$-band image from the Keck telescope. The 1.3 mm continuum emission from ALMA is presented in green contours. \label{fig4}}
\end{figure}

\clearpage







\clearpage

\begin{deluxetable}{ccrrrrr}
\tabletypesize{\scriptsize}
\tablecaption{Observed Molecular Lines and Continuum Emission\label{tbl-1}}
\tablewidth{0pt}
\tablehead{
\colhead{Species/Continuum} & \colhead{Transition} & \colhead{Frequency} & \colhead{$E_u$/$k$} & \multicolumn{2}{c}{Beam size}  & \colhead{P.A.} \\
 & & \footnotesize[GHz]  & \footnotesize[K] &  \footnotesize[arcsec] & \footnotesize[pc]\tablenotemark{a} &\footnotesize[deg]\\
}
\startdata
CS & $J$=5--4 & 244.93556 & 35.3 & 0.78$\times$0.59 & 0.0059$\times$0.0045 & 101.0 \\
$^{13}$CS & $J$=5--4 & 231.22069 & 33.3 & 0.85$\times$0.64 & 0.0064$\times$0.0048 & -72.7 \\
HCOOCH$_3$ & $J_{K_a,K_c}$=$21_{3,19}$--$20_{3,18}$ $E$ & 247.04415 & 139.9 & 0.89$\times$0.61 & 0.0067$\times$0.0046 & -68.8 \\
HCOOCH$_3$ & $J_{K_a,K_c}$=$21_{3,19}$--$20_{3,18}$ $A$ & 247.05350 & 139.9 & 0.89$\times$0.61 & 0.0067$\times$0.0046 &-68.8 \\
HCOOCH$_3$ & $J_{K_a,K_c}$=$20_{9,12}$--$19_{9,11}$ $A$ & 247.05726 & 177.8 & 0.89$\times$0.61 & 0.0067$\times$0.0046 &-68.8 \\
HCOOCH$_3$ & $J_{K_a,K_c}$=$20_{9,11}$--$19_{9,10}$ $A$ & 247.05774 & 177.8 & 0.89$\times$0.61 & 0.0067$\times$0.0046 &-68.8 \\
SiO & $J$=6--5 & 260.51802 & 43.8 & 0.80$\times$0.60 & 0.0061$\times$0.0045 & 112.1 \\
CH$_3$OH & $J_K$=$5_0$--$4_0$ $A^+$ & 241.79143 & 34.8 & 0.79$\times$0.60 & 0.0060$\times$0.0045 & 101.1 \\
CH$_3$OH & $J_K$=$10_2$--$9_3$ $A^-$ & 231.28110 & 165.3 & 0.85$\times$0.64 & 0.0064$\times$0.0048 & -72.7 \\
CH$_3$CH$_2$CN & $J_{K_a,K_c}$=$27_{0,27}$--$26_{1,26}$ & 231.31230 & 157.7 & 0.85$\times$0.64 & 0.0064$\times$0.0048 & -72.7 \\
CH$_3$CH$_2$CN & $J_{K_a,K_c}$=$24_{2,23}$--$23_{1,22}$ & 231.31324 & 132.4 & 0.85$\times$0.64 & 0.0064$\times$0.0048 & -72.7 \\
1.3 mm Continuum & $\dots$ & $\dots$ & $\dots$ & 0.82$\times$0.61 & 0.0062$\times$0.0046 & 110.4 \\
\enddata
\tablenotetext{a}{Physical scale with assuming the distance of 1.56 kpc.}

\end{deluxetable}


\clearpage




\end{document}